\documentclass[11pt]{article}

\usepackage{amsmath}
\usepackage{amssymb}
\usepackage{times}
\usepackage{dsfont}
\usepackage{theorem}
\usepackage{slashbox}

\def\=d{\,{\buildrel\rm def\over =}\,}

\title{The finite infinite range Heisenberg model\\
and microcanonical black hole statistics}
\author{Andreas Walter Aste$^{a,b}$\\
$\quad$\\
$^{a}$\emph{Department of Physics, University of Basel, 4056 Basel, Switzerland}\\
$^{b}$\emph{Paul Scherrer Institute, 5232 Villigen PSI, Switzerland}}
\date{January 21, 2016}

\begin{document}
\maketitle

\begin{abstract}
\noindent The Gelfand pattern of the reduction of the $N$-fold tensor product of the fundamental representation of the special unitary group
SU(2) by itself is studied in the framework of a finite Heisenberg model with infinite range, where $N$ spins couple to each other
with the same strength. The present findings are related to the microstatistics of non-rotating black holes for illustrative purposes.
 \\

\vskip 0.1 cm \noindent {\bf Physics and Astronomy Classification Scheme PACS (2010).}
02.20.Qs - General properties, structure, and representation of Lie groups,
04.70.-s - Physics of black holes,
04.70.Dy - Quantum aspects of black holes, evaporation, thermodynamics,
05.50.+q - Lattice theory and statistics,
97.60.Lf - Black holes
\vskip 0.1 cm \noindent {\bf Keywords.} Heisenberg model, quantum Ising model, Lie groups, group theory, tensor products, lattice theory, black holes.
\end{abstract}

%MSC
%15A69 Multilinear algebra, tensor products
%20C30 Representations of finite symmetric groups

%************************************************************************

\section*{Introduction}
\noindent Qubits are the elements of the lowest-dimensional non-trivial quantum mechanical two-state space,
and from a reductionist point of view one may argue that complex systems should be understandable
as compositions obtained from direct products and superpositions of qubit product states.\\

\noindent The special unitary group SU(2) is unusual among the matrix Lie groups in that
in the two-fold tensor product decomposition an irreducible representation (irrep) appears at most once.
For n-fold tensor products with $n \ge 3$, multiplicity does appear, and a fundamental issue is how to deal
with the repeated appearance of the same irrep \cite{Klink}. Of course, the spin group SU(2) acting on an abstract
qubit space is not necessarily related to rotational real space degrees of freedom, as the notation below may suggest.\\

\noindent In this paper, the infinite range Heisenberg model \cite{Bethe} or 'quantum Ising model' with a finite number of spins
in a homogeneous external field $B$ is investigated,
where each of the $N$ spins interacts with every other spin with equal strength given by an antiferromagnetic
renormalizable coupling parameter $g(N)>0$ depending in a non-local manner on the size of the system.
The Hamiltonian is given by
\begin{equation}
H_N = -B \sum \limits_{i=1}^{N} \frac{\sigma_3^i}{2} + g(N)\sum \limits_{i=1}^{N}
\sum \limits_{j=1}^{N} \biggl( \frac{\sigma^i_1}{2} \frac{\sigma^j_1}{2} +  \frac{\sigma^i_2}{2} \frac{\sigma^j_2}{2}+
\frac{\sigma^i_3}{2} \frac{\sigma^j_3}{2}
\biggr)  \, ,
\end{equation}
where the spin configuration $ | S \rangle$ the Hamiltonian acts upon  is an element of the $2^N$-dimensional
N-fold tensor product of the complex two-dimensional
qubit space $\mathds{C}^2_{\mathds{C}}$
\begin{equation}
\mathcal{H} = ({\mathds{C}^2_\mathds{C}})^{\otimes N} \cong \mathds{C}^{2N}_\mathds{C} \, ,
\end{equation}
since a single spin-$\frac{1}{2}$ state $| s \rangle $ is a ray in the Hilbert space $\mathds{C}^2_{\mathds{C}}$ and can be represented
by a linear combination of orthogonal basis states
\begin{equation}
| s \rangle  = u \,  | \! \uparrow \rangle + d \,  | \! \downarrow \rangle  = u 
\left(\begin{array}{c}
1 \\
0
\end{array}\right) + d
\left(\begin{array}{c}
0 \\
1
\end{array}\right) \, \quad u,d \in \mathds{C} \, , \quad
|u|^2 + |d|^2 =1 \, ,
\end{equation}
transforming under the fundamental representation $\bigl[ \frac{1}{2} \bigr]$ of SU(2) in two complex dimensions.
The Pauli matrices $\sigma_{1,2,3}$ and the spin-$\frac{1}{2}$ operators $\vec{s}=(s_1,s_2,s_3)$, $s_{1,2,3}= \frac{1}{2} \sigma_{1,2,3}$,
in a single qubit sector are given as usual by
\begin{equation}
\sigma_j =
\left(\begin{array}{cc}
\delta_{j3} & \delta_{j1}- \rm{i} \delta_{j2} \\
\delta_{j1} +  \rm{i} \delta_{j2} & - \delta_{j3}
\end{array}\right) \, , \quad \rm{i}^2=-1 \, .
\end{equation}
Since the square of total angular momentum operator $\vec{J}=\vec{s}^{\, 1} + \ldots + \vec{s}^{\, N}$ is given by
\begin{equation}
\vec{J}^{\, 2}=\Biggl( \sum \limits_{i=1}^{N} \vec{s}^{\, i} \Biggr)^2 =
\sum \limits_{i=1}^N \sum \limits_{j=1}^N \vec{s}^{\, i} \vec{s}^{\, j} \, ,
\end{equation}
the Hamiltonian can be written in the form
\begin{equation}
H_N = -B J_3 +  g(N) \vec{J}^{\, 2} \, ,
\end{equation}
and the energy eigenvalues are given therefore by
\begin{equation}
E_{J,m}=-Bm+g(N) J(J+1) \, , \quad m=-J,-J+1,\ldots +J \, , \quad 0 \le J= \frac{N}{2}, \frac{N}{2}-1, \ldots \, .
\end{equation}
The level-splitting field $B$ will not play any further r\^ole in the following.

\section*{Reduction of the N-fold Kronecker product}

In order to find degeneracy of the energy eigenvalues in the absence of an external field, one has to decompose $\mathcal{H}$
into the SU(2)-invariant subspaces according to the Kronecker (tensor, or direct) product reduction with multiplicities $\alpha_{N,K}$
\begin{equation}
\bigotimes \! ^N \biggl[ \frac {1}{2} \biggr] =
\underbrace{ \biggl[ \frac{1}{2}\biggr]  \otimes \biggl[ \frac{1}{2} \biggr]  \otimes \ldots \otimes 
\biggl[  \frac{1}{2} \biggr]  }_{N \, \, fundamental \, \, irreps} =
\bigoplus\limits_{K=0}^{N} \alpha_{N,K} \biggr[  \frac{K}{2} \biggr] \, .
\end{equation}

\noindent The following table displays the non-vanishing multiplicities of the irreps $\bigr[ \frac{K}{2} \bigr] $ for $N=1,\ldots 12$
with $J=K/2$\\

\def\arraystretch{1.2}
\begin{tabular}{l|ccccccccccccccc}
\backslashbox[10pt][c]{$N$}{$J$} & $0$ & $\frac{1}{2}$ & $1$ & $\frac{3}{2}$ & $2$ & $\frac{5}{2}$ & 3 &
$\frac{7}{2}$ & $4$ & $\frac{9}{2}$ & $5$ & $\frac{11}{2}$ & 6 &  \\\hline
$1$     &          &  $1$   &         &        &        & & &&&&&&\\
$2$     &  $1$   &          & $1$    &        &         & & &&&&&&\\
$3$     &  &  $2$  & & $1$          & & &&&&&&\\
$4$     & $2$   &  & $3$ &   & $1$          & & &&&&&&\\
$5$     &  & $5$  &  & $4$  &   & $1$   & &&&&&&\\
$6$     & $5$   &  & $9$  &  & $5$ &  & $1$  &      &&&&&\\
$7$     &         & $14$ & & $14$  &  & $6$ &  & $1$   &&&&&\\
$8$     & $14$  &  & $28$  &  & $20$ &  &  $7$   &    &  $1$     &&&\\
$9$     &   & $42$  &  & $48$  &  & $27$ &  &  $8$   &    &  $1$ &&    &\\
$10$    & $42$  &  & $90$  &  & $75$ &  &  $35$   &    &  $9$ & & $1$      &\\
$11$     & & $132$ &   & $165$  &   & $110$  && $44$ && $10$ && $1$ \\
$12$    & $132$ &   & $297$  &   & $275$  && $154$ && $54$ && $11$ && $1$ 
\end{tabular} \\
$\quad$\\

\noindent For $N=2$, the product $\bigl[ \frac{1}{2} \bigr] \otimes \bigl[ \frac{1}{2} \bigr]$ decays into
one singlet $\bigl [0 \bigr]$ and one triplett $\bigl[ 1 \bigr]$, for $N=3$ the product $\bigl[ \frac{1}{2} \bigr]
 \otimes \bigl[ \frac{1}{2} \bigr] \otimes \bigl[ \frac{1}{2} \bigr]$ results in
\begin{equation}
\biggl[ \frac{1}{2} \biggr] \otimes \biggl[ \frac{1}{2} \biggr] \otimes \biggl[ \frac{1}{2} \biggr] =
\biggl[ \frac{1}{2} \biggr] \otimes \biggl(  \biggl [0 \biggr] \oplus \biggl[ 1 \biggr] \biggr) =
2 \biggl[ \frac{1}{2} \biggr] \oplus \biggl[ \frac{3}{2} \biggr] \, ,
\end{equation} 
and it is clear from the well-known Clebsch-Gordan decomposition of the Kronecker product
of two SU(2)- irreps of dimension $2j_1+1$ and $2j_2+1$ corresponding to angular momenta $j_1$,$j_2$, respectively
\begin{equation}
[j_1] \otimes [j_2] = \bigoplus \limits_{j=|j_1-j_2|}^{j_1+j_2}  [j] \, ,
\end{equation}
how two construct the table inductively, however, there is no simple construction scheme
for an explicit expression.
Still, a thorough study of the pattern leads to the compact formula for the non-vanishing multiplicities
(here with $J=K/2$), which can be proved by induction
\begin{equation}
\alpha_{N,K} = \frac{2J+1}{ N+1} \frac{(N+1)!}{(N/2-J)!(N/2+J+1)!} \, , \quad
J=
 \left\{ \begin{array}{rcc}
0, 1, \ldots \frac{N}{2} & : &  N \, \, even  \\
\frac{1}{2}, \frac{3}{2}, \ldots \frac{N}{2} & : &  N \, \, odd \end{array} \right. \, .
\end{equation}
Of course one has for the dimension of the full Hilbert space $\mathcal{H}$
\begin{equation}
\sum \limits_{K} (K+1) \alpha_{N,K} = 2^N \, , \label{sum}
\end{equation}
where
\begin{equation}
(K+1) \alpha_{N,K} = \frac{  (2J+1)^2 N!}{(N/2-J)! (N/2+J+1)!} = \frac{(2J+1)^2}{N+1} { N+1 \choose N/2-J } \, .
\label{degen}
\end{equation}

\section*{Large N, J limit}
For large values of N and values of $J$ such that the degeneracy eq. (\ref{degen}) is comparatively large,
an approximation of binomial coefficients based on their relation to the normal distribution is
\begin{equation}
{ n \choose k } \simeq \frac{2^{n+1/2}}{\sqrt{n \pi}} e^{-\frac{(k-n/2)^2}{(n/2)}} \, .
\end{equation}
From this expression one derives the asymptotic formulae
\begin{displaymath}
(K+1) \alpha_{N,K} \simeq (2J+1)^2 \frac{2^{N+3/2}}{\sqrt{(N+1)^3 \pi}} e^{- 2 (J+1/2)^2/(N+1)}
\end{displaymath}
\begin{equation}
\simeq \frac{2^N}{\sqrt{2^{-7} N^3 \pi}} J(J+1) e^{- 2 J(J+1)/N}
\simeq  \frac{2^N}{\sqrt{2^{-7} N^3 \pi}} J^2e^{- 2 J^2/N}
\end{equation}
Considering $J$ as a continuous parameter for the moment, the discrete sum rule eq. (\ref{sum}) becomes
\begin{equation}
\int \limits_{0}^\infty dJ \,   \frac{2^N}{\sqrt{2^{-7} N^3 \pi}} J^2 e^{- 2 J^2/N} = 2^N
\end{equation}
as a rescaled version of the Gaussian identity
\begin{equation}
\int \limits_{0}^\infty dx \, x^2 e^{-x^2}= \frac{\sqrt{\pi}}{4} \, .
\end{equation}
The most abundant irreps are located around
\begin{equation}
J \simeq \sqrt{N/2} \, . \label{JN}
\end{equation}
For large $N$, $\mbox{erf}(\sqrt{8})-4 \sqrt{2/\pi}e^{-8} \simeq 99.8866\%$
of all states are located in the interval $0 < J < 2 \sqrt{N}$, however, the width of the distribution is not sharp and also
of the order of $\sqrt{N/2}$.
From the expressions above, all relevant thermodynamic properties can be calculated in the case of
large systems potentially using Hubbard-Stratonovich transformation techniques \cite{Stratonovich,Hubbard}.

\section*{Microcanonical black hole ensemble}
The literature contains a plethora of more or less naive attempts to derive expressions for
the thermodynamic behavior of black holes \cite{Aste1,Aste2} by counting microscopic degrees of freedom
of many kinds.
The assumption that the horizon of a spherically symmetric black hole of mass $M$ and corresponding area $A=4 \pi R_S^2 =
16 \pi G^2 M^2/c^4$ ($G$ is the gravitational constant, $c$ the speed of light in vacuo, and
$R_S = 2 GM/c^2$ the Schwarzschild radius)  is segmented into area quanta $a= 4 \ln (2) l_P^2$
of the order of the Planck length squared $l_P^2 = \hbar G/c^3$, inspired Bekenstein, Mukhanov \cite{Mukhanov}
and Kastrup \cite{Kastrup} to postulate and study discrete energys levels
for such  black holes in the spirit of the Bohr-Sommerfeld quantization in atomic physics,
with a mass spectrum $M_n = \mu \sqrt{n}$, $n \in \mathds{N}_0$, and an energy level degeneracy of the
form $\nu(n)=b^n$ with $\mu= \sigma m_P = \sigma \sqrt{\hbar c/G}$, $\sigma=o(1)$, $b>1$.\\

\noindent Inspired by the Heisenberg spin picture above one may consider the special case with energy spectrum
\begin{equation}
E(n)=\sqrt{\frac{\ln(2)}{4 \pi}} E_P \sqrt{n} \, , \quad n=0,1,2,3,\ldots \,  , \quad E_P= \sqrt{ \hbar c^5 / G} \, ,
\label{quaint}
\end{equation}
with a degeneracy given by
\begin{equation}
\nu(n) = 2^n \, .
\end{equation}
The microcanonical partition function $\Omega(E)$ is the number of energy eigenstates with an energy below
$E$
\begin{equation}
\Omega(E) = \sum \limits_{n=0}^{n(E)-1} 2^n = 2^{n(E)} = 2^{\frac{4 \pi}{\ln(2)} \bigl( \frac{E}{E_P} \bigr)^2}
= e^{4 \pi \bigl( \frac{E}{E_P} \bigr)^2} \, .
\end{equation}
The microcanonical temperature becomes 
\begin{equation}
k_B T = \frac{\Omega}{\omega} \, , \quad \omega= \frac{ \partial \Omega}{\partial E} =
8 \pi \frac{E}{E_P^2} \Omega \, ,
\end{equation}
hence one arrives at the Bekenstein-Hawking temperature
\begin{equation}
k_B T = \frac{E_P^2}{8 \pi M c^2} \, ,
\end{equation}
which is quantized if the model is taken seriously,
and the entropy is
\begin{equation}
S= k_B \ln \Omega = k_B \frac{A}{4 l_P^2} \, .
\end{equation}
Of course, it is possible to assume that the idealized energy spectrum of a spherically symmetric black hole
alone in an infinite asymptotically flat universe is non-degenerate and therefore
\begin{equation}
n(E)= e^{4 \pi \bigl( \frac{E}{E_P} \bigr)^2}  \, ,
\end{equation}
or
\begin{equation}
E(n)=E_P \sqrt{ \frac{ \ln(n)}{4 \pi}} \, .  \label{smooth}
\end{equation}
Of course, an unstable object is never isolated since it quantum superposes itself with its decay states,
thereby blurring its energy spectrum into the complex plane.
Even if the ansatz eq. ( \ref{quaint}) is not valid, it remains tempting to figure the black hole horizon
as a network of $N$ area quanta interacting (locally) as spins with decreasing coupling strenght
$g(N)$ as $N$ increases together with a decreasing surface gravity $\sim R_S^{-1}$, interpolating the
discrete spectrum eq. (\ref{quaint}) to something smoother, closer in style to eq. (\ref{smooth}).
Of course, the exactly solvable Heisenberg model on a complete graph probably is to coarse to display
a realistic spectrum adequately.
Still, assuming that the number of spins is given by the surface or the energy of the black hole
\begin{equation}
N=\frac{4 \pi R_S^2}{4 \ln(2) l_P^2} = \frac{4 \pi G}{ \ln(2) \hbar c^5} E^2
\label{area_energy}
\end{equation}
together with the Ansatz
\begin{equation}
g(N)= \sqrt{\frac{\ln(2)}{\pi}} \frac{E_P}{\sqrt{N}} \, ,
\end{equation}
the collapse of non-rotating matter with energy $E$ will lead to the creation of $N$ qubits
coupled to a total qubit momentum $J \simeq \sqrt{N/2}$,
such that
\begin{equation}
E=g(N) J^2 =  \sqrt{\frac{\ln(2)}{\pi}} \frac{E_P}{\sqrt{N}} \frac{N}{2} = \sqrt{\frac{\ln(2)}{4 \pi}} E_P \sqrt{N}
\end{equation}
in accordance with eqns. (\ref{area_energy}) and  (\ref{JN}). Such a size of $J$ corresponds to the
radius of the black hole as a fuzzy sphere.

\section*{Conclusions}
A group theoretical result concerning the n-fold tensor product of the defining, two-dimensional pseudo-real representation
of SU(2) has been presented, including the calculation of the spectrum of the finite infinite range Heisenberg model as a new result.
The present letter is a pedagogical illustration of some aspects of current theories inspired by string or loop quantum gravity approaches
aiming at the construction of quantum space time models.

\end{document}